\documentclass{article}
\usepackage[preprint]{spconf,amsmath,graphicx}
\usepackage{multirow}
\usepackage{subfig}
\usepackage{cite}
\usepackage{tikz}

\title{Light-Field View Synthesis using A Convolutional Block Attention Module}
\name{M. Shahzeb Khan Gul$^{\dagger}$, M. Umair Mukati$^{\star}$, Michel B{\"a}tz$^{\dagger}$, S{\o}ren Forchhammer$^{\star}$, Joachim Keinert$^{\dagger}$\sthanks{Published in the IEEE 2021 International Conference on Image Processing (IEEE ICIP 2021), scheduled for 19-22 September 2021 in Anchorage, Alaska, United States. Personal use of this material is permitted. However, permission to reprint/republish this material for advertising or promotional purposes or for creating new collective works for resale or redistribution to servers or lists, or to reuse any copyrighted component of this work in other works, must be obtained from the IEEE. Contact: Manager, Copyrights and Permissions / IEEE Service Center / 445 Hoes Lane / P.O. Box 1331 / Piscataway, NJ 08855-1331, USA. Telephone: + Intl. 908-562-3966.}} 
\address{$^{\dagger}$ Moving Picture Technologies, Fraunhofer IIS, 91058, Erlangen, Germany \\$^{\star}$ DTU Fotonik, Technical University of Denmark, {\O}rsteds Plads, Kgs. Lyngby - 2800, Denmark}

\begin{document}
\copyrightnotice{\copyright\ IEEE 2021. }
\ninept
\maketitle
\begin{abstract}
Consumer light-field (LF) cameras suffer from a low or limited resolution because of the angular-spatial trade-off. To alleviate this drawback, we propose a novel learning-based approach utilizing attention mechanism to synthesize novel views of a light-field image using a sparse set of input views (i.e., 4 corner views) from a camera array. In the proposed method, we divide the process into three stages, stereo-feature extraction, disparity estimation, and final image refinement. We use three sequential convolutional neural networks for each stage. A residual convolutional block attention module (CBAM) is employed for final adaptive image refinement. Attention modules are helpful in learning and focusing more on the important features of the image and are thus sequentially applied in the channel and spatial dimensions. Experimental results show the robustness of the proposed method. Our proposed network outperforms the state-of-the-art learning-based light-field view synthesis methods on two challenging real-world datasets by 0.5 dB on average. Furthermore, we provide an ablation study to substantiate our findings.
\end{abstract}
\begin{keywords}
Light-field, View synthesis, Deep-learning, 
\end{keywords}

\section{Introduction}
\label{sec:intro}
Traditional cameras capture only the intensities of the incident light-rays at the photosensor, whereas, light-fields collect massive information of the scene by additionally acquiring the directional information of the incident light-rays. This additional angular information becomes beneficial in many applications such as depth estimation, post-capture refocusing, and 3D reconstruction. 
 
Light-fields can be captured through many different ways. Conventional methods include camera arrays \cite{wilburn2005high}, which can capture a LF simultaneously in time but have a limited angular resolution because of the numbers of cameras and the necessary spacing. Gantry systems \cite{gul2020high}, which sample a high angular resolution light-field by sequentially capturing each viewpoint, cannot capture dynamic content. On the other hand, hand-held light-field cameras, such as from Lytro and Raytrix \cite{perwa2018raytrix}, offer a feasible solution for light-field capture, based on the plenoptic camera designs \cite{ng2005light}. However, it suffers from the spatial and angular resolution trade-off due to the sensor limitation.

In recent years, different light-field view synthesis approaches have been published aiming to solve the limited angular resolution problem. The success of deep learning compared to traditional methods for numerous computer vision and image processing problems, such as multi-view stereo \cite{poggi2019guided, gul2019pixel}, optical flow \cite{liu2019selflow}, and super-resolution \cite{kohler2019toward}, inspired the researchers to develop learning-based view interpolation methods \cite{kalantari2016learning, gul2018spatial, navarro2019learning, liu2020dcm, wing2018fast, wang2018end}. These methods can be broadly divided into two categories: image-based rendering (IBR) and depth image-based rendering (DIBR) methods. The IBR methods synthesizing the novel views without explicitly modeling the scene depth \cite{wing2018fast, wang2018end}, tend to fail with increasing baseline of the input views, whereas, depth-based methods can handle inputs with wider baseline (i.e. large disparities) \cite{navarro2019learning, yue2020benchmark}.

In this paper, we demonstrate for the first time that CBAM used in a different context, i.e., object detection, can be used for light-field view synthesis. We divide the whole process into three stages, i.e., stereo-feature extraction, disparity estimation, and final image refinement. In the first step, we extract features directly from stereo pairs and we feed these stereo features into a disparity estimation network to estimate target disparity maps corresponding each input view. Multiple disparity vectors per target view alleviate the interpolation error due to the warping of input images. In the end, an image refinement network using a convolutional block attention module takes the warped images as input, and outputs the final novel target view. Attention modules have been studied and used to a great extent in the literature \cite{wang2017residual}, \cite{hu2018squeeze}, \cite{woo2018cbam}. Attention mechanisms not only guide the network to focus on important features, but they also improve the representation of the regions of interest. 
\begin{figure*}[!]
\centering
\includegraphics[width=0.8\linewidth]{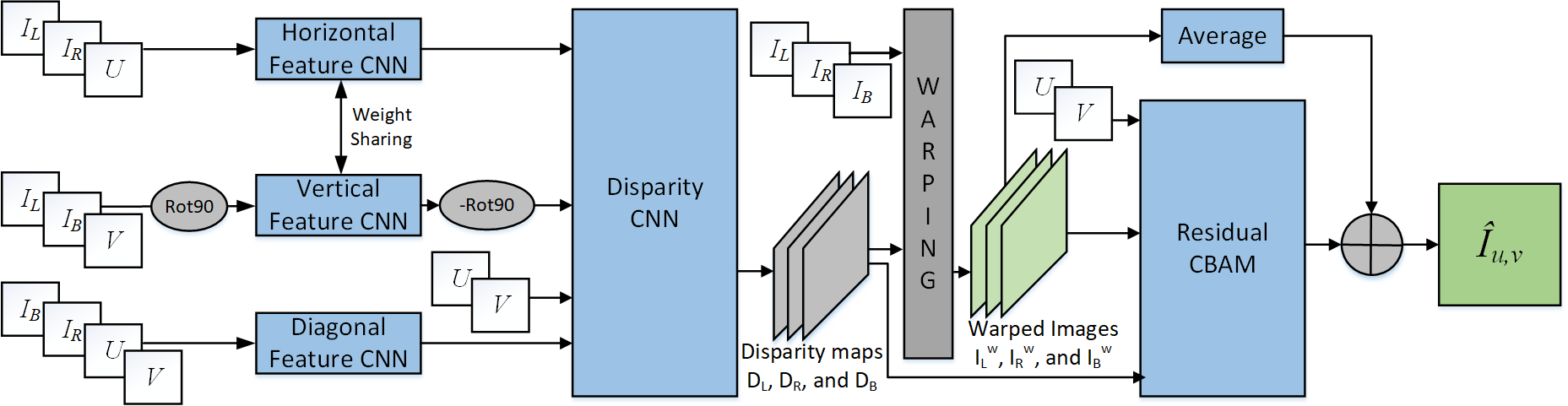}
\setlength{\belowcaptionskip}{-15pt} 
\caption{Overall flow of the proposed network. Here, $I_L$, $I_R$, and $I_B$ are the selected input images as shown in Fig. \ref{view_select}. $U$ and $V$ are of the size of the input views and repeatedly contain the $u$ and $v$ coordinate of the target view to be rendered, respectively.}
\label{flow}
\end{figure*}

The remainder of the paper is structured as follows. In Section~\ref{sec:literature}, we provide a comprehensive literature review on state-of-the-art view synthesis algorithms. Section~\ref{sec:method} explains the proposed network architecture. Experimental results are then presented and discussed in Section~\ref{sec:exp} before the paper is concluded in Section~\ref{conc}.
\section{Related Work}
\label{sec:literature}
\subsection{Image-based rendering (IBR) method}
In literature, some methods approach the light-field view synthesis problem as a specific case of signal reconstruction. They exploit the intrinsic sparsity of the light-field images. In \cite{shi2014light}, a captured light-field is considered to be sparse in the continuous Fourier domain; hence, novel views can be reconstructed from a small number of 1D viewpoint trajectories. Levin and Durand \cite{levin2010linear} synthesize the novel views using 3D focal stack sequences with an assumption that the light-field contains only Lambertian surfaces. 

Many researchers used sparse coding as a tool to interpolate novel views. In \cite{marwah2013compressive}, a global dictionary is learned from light-field patches to synthesize novel views from a set of input light-field views. On the other hand, a local dictionary learned using the central region of the light-field images is successfully applied for the interpolation of novel views in \cite{schedl2015directional}, \cite{schedl2018optimized}.

In \cite{yoon2017light} and \cite{gul2018spatial}, the authors approached the light-field view synthesis problem as angular domain super-resolution. In \cite{yoon2017light}, two sequential convolutional neural networks (CNNs) take sub-aperture views as inputs to increase the spatial and angular resolution of the light-field consecutively. On the other hand, \cite{gul2018spatial} applied the CNNs directly on lenslet images to increase the resolution in both spatial and angular domains. Wang \textit{et al.} \cite{wang2018end} proposed to use stacked 3D volumes of epipolar plane images in a CNN framework. In \cite{wing2018fast}, spatio-angular alternating convolutions (i.e., pseudo 4D-CNN) are performed to incorporate 4D light-field data for novel view interpolation. On the other hand, our proposed network synthesizes the novel view by explicitly modeling the scene geometry.
\begin{figure}[t!]
\centering
\includegraphics[width=1\linewidth]{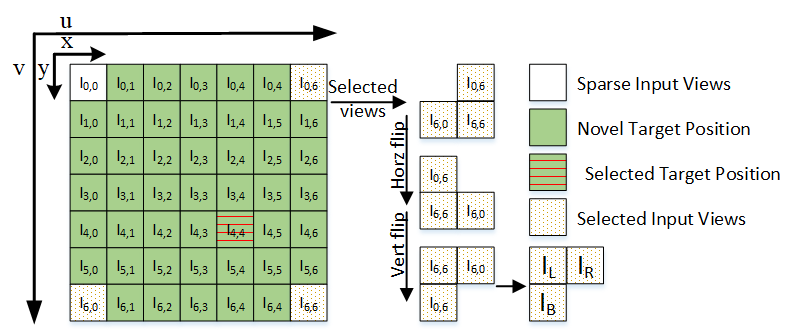}
\setlength{\belowcaptionskip}{-17pt} 
\caption{Based on the angular coordinate of the target view, we select three input views from a given sparse set of four corner views. Input views with the least absolute distance from the target view are selected. To maintain the correspondence direction constant among different input combinations, we flip the selected inputs views in horizontal or vertical direction accordingly to make an upper left triangle shape.}
\label{view_select}
\end{figure}
\subsection{Depth image-based rendering (DIBR) method}
Typically, depth-based view synthesis methods are divided into two steps. At first, these methods estimate the disparity at a target view using the sparse light-field views and then utilize it to synthesize the target view by warping of the input views. In \cite{kalantari2016learning}, two CNNs are trained to jointly estimate the depth and synthesize a view in an end-to-end fashion. In \cite{navarro2019learning}, a feature extraction CNN is introduced as the first part of the network architecture. Moreover, the methods estimates four disparity maps corresponding to each input view and then utilize a selection CNN for the selection of pixels from warped images. Jin \textit{et al.} \cite{jin2020learning} proposed CNN estimating 4D depth maps for a high angular resolution light-field, i.e., providing depth for each light ray in the 4D light-field. The resulting 4D depth is then utilized to synthesize all novel views by backward warping. As a final step, a CNN light-field blending module is employed to get the final light-field.

Recently, a view synthesis method based on multi-plane image (MPI) \cite{zhou2018stereo} was introduced, where a scene is represented by a stack of RGBA planes at different depths. In \cite{mildenhall2019local}, this idea is extended to light-field using a 3D-CNN to infer an MPI. Flynn \textit{et al.} \cite{flynn2019deepview} proposed to use the variational optimization framework in conjunction with deep learning to improve the reconstruction quality of MPIs.

While our algorithm is inspired by \cite{navarro2019learning}, it introduces a new combination of attention mechanism to efficiently deal with the occluded regions as shown in Fig. \ref{test123}.
\subsection{Attention mechanism}
Very recently, attention mechanisms have become a very popular method to enhance a deep neural network. Attention modules try to make a neural network focus on salient regions of its feature representation. In \cite{wang2017residual}, a residual attention network is used for image classification. The Squeeze-and-excitation (SE) block attention module introduced in \cite{hu2018squeeze}, improves the classification accuracy by exploiting channel inter-dependencies. In \cite{woo2018cbam}, the authors introduced a CBAM by adding a spatial attention mechanism to a SE block to validate performance of object detection. In \cite{suzuki2020residual} a variant of CBAM is used in video frame interpolation to refine the interpolated frame. Our proposed method uses the attention mechanism for a similar purpose, i.e., image refinement. We integrate CBAM as proposed in \cite{woo2018cbam} into our light-field view interpolation pipeline, using both channel and spatial attention.
\section{Proposed Method}
\label{sec:method}
Let $L(x,y,u,v)$ denote a light-field according to two-plane parametrization as shown in Fig. \ref{view_select}, where $(x, y)$ are the spatial coordinates and $(u, v)$ are the angular coordinates. Each sub-aperture view in the light-field is denoted as $I_{u,v}$. 
\subsection{Network Architecture}
The architecture of our proposed network is shown in Fig. \ref{flow}. To synthesize high-quality novel views, we divide the proposed method into three stages: stereo feature extraction, disparity estimation, and image refinement. Let $L^{S}$ be four corner views of a light-field, our proposed method makes use of the three closest corner views to reconstruct a high-quality image $I_{u_t,v_t}$ at the target position:
\begin{equation}
\hat{I}_{u_t,v_t} = \mathit{f}(I_L, I_R, I_B, u_t, v_t),
\end{equation}
where $\mathit{f}$ is the function that we aim to learn and which should estimate a view $\hat{I}_{u_t,v_t}$ as similar as possible to the ground-truth image $I_{u_t,v_t}$. $I_L$, $I_R$, and $I_B$ are the three selected corner views for the target position $(u_t, v_t)$. Out of the four, the three reference views are selected as input to the network based on their minimum distance to the target view (as depicted in Fig. \ref{view_select}). Selecting three corner views as opposed to four reduces the overall complexity of the network. To maintain constant geometry among different inputs, we flip the images left-to-right and/or up-to-down accordingly so that we always have an upper left triangle shape of the selected input views (see Fig. \ref{view_select}). Note that the flipping operation will also change the angular coordinate of the target view. Similarly, we flip back the output of the network to get the final reconstructed image.
\subsubsection{Stereo feature extraction network}
We first extract stereo features from the sparse set of selected input views. The architecture of the feature extraction module $f_e$ is based on \cite{navarro2019learning}. The network $f_e$ consists of six convolutional layers with 3$\times$3 kernels and two average pooling layers with 16$\times$16 and 8$\times$8 kernels. However, instead of extracting features for each image individually, as done in \cite{navarro2019learning}, we extract features directly from stereo pairs. We group the input views into three stereo pairs, namely horizontal ($I_L$,$I_R$), vertical ($I_L$,$I_B$), and diagonal ($I_B$,$I_R$). Horizontal and vertical stereo pairs use the same architecture for feature extraction. The reason for having a shared network here is that the geometry between the images can be made identical to the horizontal case when the vertical stereo pair is rotated by 90$^{\circ}$ counter-clockwise. Furthermore, this reduces the complexity of the network and thus simplifies the trainability. In addition to the stereo pair as input, the network also takes $U$ and/or $V$ image planes, where $U(x, y) = u_t, V(x, y) = v_t  \quad \forall (x, y)$. These image planes aid the network in collecting appropriate information from each stereo pair based on the position of the novel view.

Let $F_\mathtt{horz} = f_e(I_L, I_R, U)$, $F_\mathtt{vert} = f_e(I_L, I_B, V)$, and $F_\mathtt{diag} = f_e(I_B, I_R, U, V)$ be the computed feature volumes for the three stereo pairs. These 32-channel feature volumes are then concatenated and passed to the disparity estimation network.
\subsubsection{Disparity estimation network}
The goal of this module is to estimate a disparity map per input view valid at the target position. Multiple disparity values per target view pixel allow for a more effective handling of  occluded regions as compared to \cite{kalantari2016learning} where a single disparity map is estimated for all input views. The architecture of the disparity estimation network $f_d$ is inspired by \cite{navarro2019learning}. The network consists of seven convolutional layers with 3$\times$3 filter size. The first four layers use dilated convolutions at rates 2, 4, 8, and 16, respectively. The network takes the features extracted from the stereo pairs and angular coordinate image planes $U$ and $V$ as input and output disparity maps $D_i$, for $i \in \{L, R, B\}$.
\begin{figure}[!]
\centering
\includegraphics[width=0.65\linewidth]{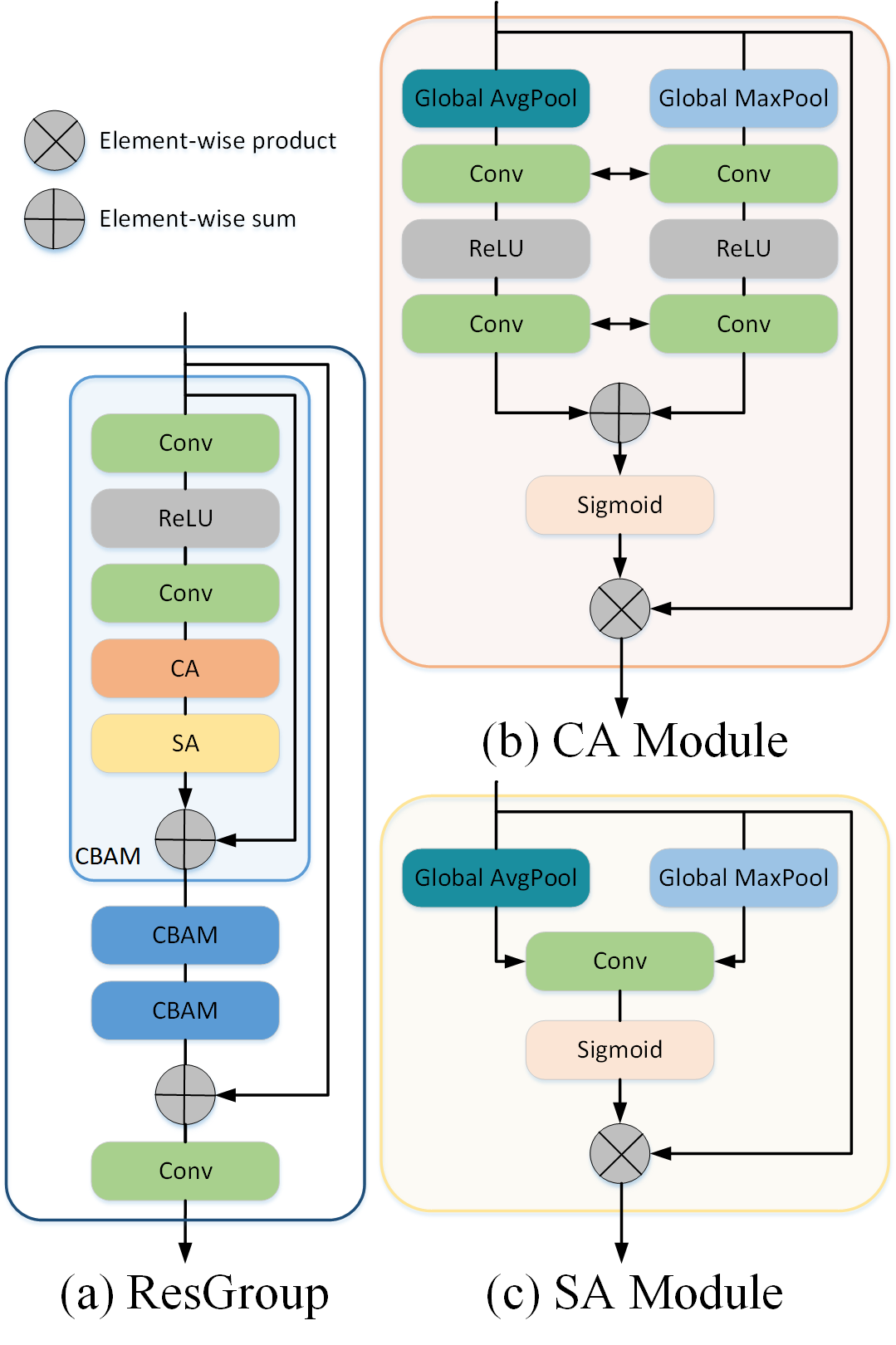}
\setlength{\belowcaptionskip}{-15pt} 
\caption{Illustration of ResGroup, CA module, and SA module.}
\label{CBAM}
\end{figure}
\subsubsection{Image refinement network}
The estimated disparity maps are used to warp each corner view in order to have them registered with the target one. After warping the input views $I^{w}_i$, for $i \in \{L, R, B\}$, at the target location using the estimated disparity maps, we estimate the final image using a residual CBAM refinement network. The architecture of the refinement network consists of a head and tail 3$\times$3 convolution layer with five residual groups in between, referred to as \textit{ResGroups} from now, each of which consists of three CBAMs \cite{woo2018cbam}. As shown in Fig. \ref{CBAM} (a), each CBAM has two 3$\times$3 convolution layers with ReLU activation in between, followed by a channel attention (CA) and spatial attention (SA) module before the residual connection. Both attention modules, channel and spatial, compute complementary attentions focusing on `what' and `where' respectively. The architecture of the CA module is shown in Fig. \ref{CBAM} (c).  We first aggregate the spatial information using average-pooling and max-pooling operations. Features descriptors from both operations are then forwarded to a shared two 1$\times$1 convolution layer network to produce a 1D channel attention map. Similarly, the architecture of the SA module is illustrated in Fig. \ref{CBAM} (b). We aggregate the channel information using two pooling operations followed by a 7$\times$7 convolution layer producing a 2D spatial attention map. The final output image at the target position is obtained by adding the output of the refinement network and the average of the warped views.
\subsection{Loss Function}
The model is trained by optimizing the sum of the following loss functions:
\begin{equation}
\small
\mathcal{L}_\mathtt{final} = \Vert{I_{u,v} - \hat{I}_{u,v}}\Vert_1 + \lambda_1\Vert{\nabla I_{u,v} - \nabla\hat{I}_{u,v}}\Vert_1,
\end{equation}
\begin{equation}
\small
\mathcal{L}_\mathtt{warp} = \lambda_2\sum_{i}\Vert{I_{u,v} - I^w_i}\Vert_1 + \lambda_3\sum_{i} \Vert{\nabla I_{u,v} - \nabla{I^w_i}}\Vert_1.
\end{equation}
$I^w_{i}$ is the warped view from one stereo pair using the estimated disparity map, where $i \in \{L, R, B\}$. $I_{u,v}$ and $\hat{I}_{u,v}$ are ground-truth and estimated novel views, respectively. Moreover, we experimentally set $\lambda_1 = 0.5$, $\lambda_2 = 0.25$, and $\lambda_3 = 0.125$. The first term in both losses is the $L_1$ loss between the predicted and the ground-truth images, whereas, the second term preserves texture by taking $L_1$-norm between spatial gradients.

\subsection{Training}
The proposed model is implemented using TensorFlow as framework. The model is trained on the light-field dataset \textit{Flowers} presented in \cite{srinivasan2017learning1}. This dataset consists of 3343 images of flowers. We randomly select 100 images for testing and used the remaining images for training the network. Each light-field in the dataset has a spatial resolution of 540$\times$372 and an angular resolution 14$\times$14. During the experiment, we only consider the center 7$\times$7 grid of views because the corner views are affected by vignetting. As a data augmentation, we first randomly crop inputs to a size of 192$\times$192 then randomly perform gamma correction using a value from the range [0.4, 1]. The pixel values are normalized to a range of [-1, 1]. The optimization of the proposed network is done with the ADAM  optimizer where $\beta_1$ and $\beta_2$ are set to 0.9 and 0.99. The learning rate is set to 0.0001, while the batch size is 8. The model converges after 1000 epochs, and it takes approximately 2.5 days to train on a GeForce GTX 1080 Ti GPU.
\section{Experimental Results}
\label{sec:exp}
In this section, we compare the proposed method with state-of-the-art methods both quantitatively and qualitatively. We evaluate all the methods on two test datasets, one contains 100 light-fields of flowers that we randomly select from \cite{srinivasan2017learning1}. The other test set \textit{Diverse} consists of 30 light-fields \cite{kalantari2016learning} -- mostly outdoor scenes. Besides, we also conducted ablation studies to assess the effectiveness of the various components of our network.
\begin{table}[t!]
\centering
\footnotesize
\setlength\tabcolsep{7pt}
\begin{tabular}{c|cc|cc}
\multicolumn{1}{c|}{\begin{tabular}[c]{@{}c@{}}\end{tabular}} & \multicolumn{2}{c|}{\textit{Diverse} \cite{kalantari2016learning}}  & \multicolumn{2}{c}{\textit{Flowers} \cite{srinivasan2017learning1}}                                                                                                                                                                                                   \\ 
Method            & \begin{tabular}[c]{@{}c@{}}PSNR\\ (dB)\end{tabular} & \begin{tabular}[c]{@{}c@{}}MS-SSIM\\ (dB)\end{tabular} & \begin{tabular}[c]{@{}c@{}}PSNR\\ (dB)\end{tabular} & \begin{tabular}[c]{@{}c@{}}MS-SSIM\\ (dB)\end{tabular} \\ \hline
Kalantari \textit{et al.} \cite{kalantari2016learning} & 39.56 & 24.31 & 40.22 & 22.47\\ 
Navarro \textit{et al.} \cite{navarro2019learning} & 40.75 & 25.57 & 42.67 & 25.04\\ 
LFVS (ours) & 41.14 & 25.89 & 43.13 & 25.44\\
LFVS-AM (ours)   & \textbf{41.33} & \textbf{26.05} & \textbf{43.17} & \textbf{25.54}\\ 
\end{tabular}
\setlength{\belowcaptionskip}{-8pt} 
\caption{Quantitative results on the test datasets \textit{Flowers} and \textit{Diverse}. For each row, average PSNR (dB) and MS-SSIM (dB) values for Y channel of the image achieved on all intermediate rendering positions are listed for all methods.}
\label{psnr}
\end{table}
\begin{figure}[!]
\captionsetup[subfloat]{farskip=2.5pt,captionskip=1.5pt}
\captionsetup[subfloat]{font=scriptsize}
\centering
%
%
%
%
\begin{tikzpicture}
  \node[anchor=south west,inner sep=0] (image) at (0,0)  
  {\includegraphics[width=0.19\linewidth]{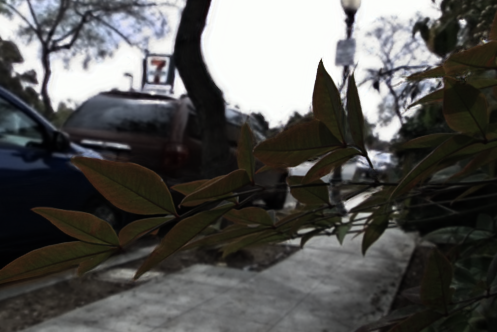}};
  \begin{scope}[x={(image.south east)},y={(image.north west)}]
        \draw[red,thick] (0.62,0.7) rectangle (0.75,0.82);
    \end{scope} 
  \end{tikzpicture}\hfill
\begin{tikzpicture}
  \node[anchor=south west,inner sep=0] (image) at (0,0)
  {\includegraphics[width=0.19\linewidth]{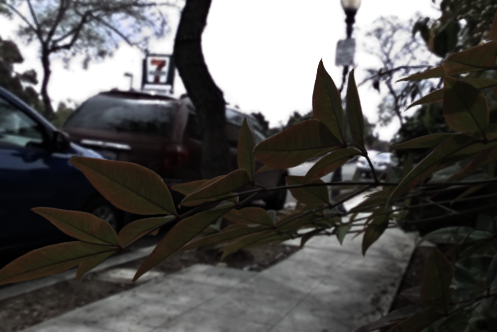}};
  \begin{scope}[x={(image.south east)},y={(image.north west)}]
        \draw[red,thick] (0.62,0.7) rectangle (0.75,0.82);
    \end{scope} 
  \end{tikzpicture}\hfill
\begin{tikzpicture}
  \node[anchor=south west,inner sep=0] (image) at (0,0)
  {\includegraphics[width=0.19\linewidth]{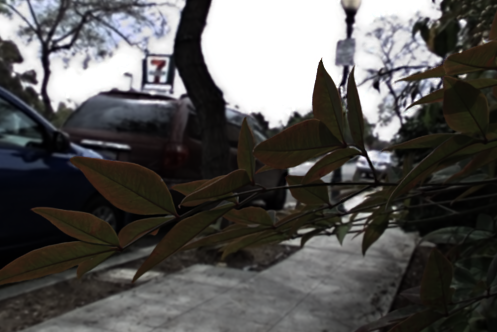}};
  \begin{scope}[x={(image.south east)},y={(image.north west)}]
        \draw[red,thick] (0.62,0.7) rectangle (0.75,0.82);
    \end{scope} 
  \end{tikzpicture}\hfill
\begin{tikzpicture}
  \node[anchor=south west,inner sep=0] (image) at (0,0)
  {\includegraphics[width=0.19\linewidth]{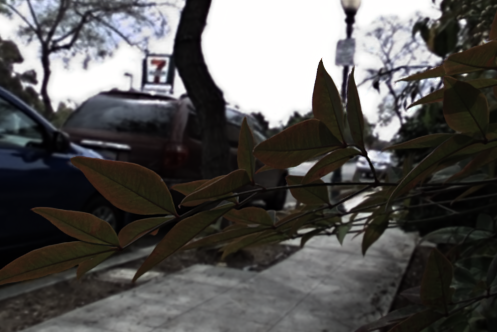}};
  \begin{scope}[x={(image.south east)},y={(image.north west)}]
        \draw[red,thick] (0.62,0.7) rectangle (0.75,0.82);
    \end{scope} 
  \end{tikzpicture}\hfill
\begin{tikzpicture}
  \node[anchor=south west,inner sep=0] (image) at (0,0)
  {\includegraphics[width=0.19\linewidth]{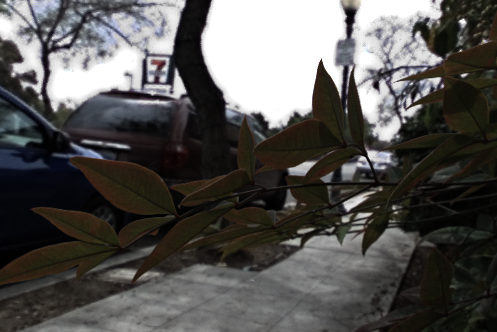}};
  \begin{scope}[x={(image.south east)},y={(image.north west)}]
        \draw[red,thick] (0.62,0.7) rectangle (0.75,0.82);
    \end{scope} 
  \end{tikzpicture}

  \subfloat[29.28 dB\label{fig:test11}]
  {\includegraphics[trim=308 232 125 60, clip,width=0.19\linewidth]{Images/kalantari1.png}}\hfill
  \subfloat[30.37 dB\label{fig:test12}]
  {\includegraphics[trim=308 232 125 60, clip,width=0.19\linewidth]{Images/navarro1.png}}\hfill
  \subfloat[31.75 dB\label{fig:test13}]
  {\includegraphics[trim=308 232 125 60, clip,width=0.19\linewidth]{Images/proposed_wa1.png}}\hfill
  \subfloat[33.46 dB\label{fig:test14}]
  {\includegraphics[trim=308 232 125 60, clip,width=0.19\linewidth]{Images/proposed1.png}}\hfill
  \subfloat[\label{fig:test15}]
  {\includegraphics[trim=308 232 125 60, clip,width=0.19\linewidth]{Images/1528.png}}

  \begin{tikzpicture}
  \node[anchor=south west,inner sep=0] (image) at (0,0)  
  {\includegraphics[width=0.19\linewidth]{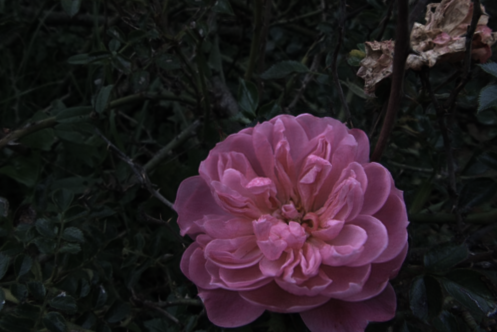}};
  \begin{scope}[x={(image.south east)},y={(image.north west)}]
        \draw[red,thick] (0.78,0.8) rectangle (0.88,0.9);
    \end{scope} 
  \end{tikzpicture}\hfill
\begin{tikzpicture}
  \node[anchor=south west,inner sep=0] (image) at (0,0)
  {\includegraphics[width=0.19\linewidth]{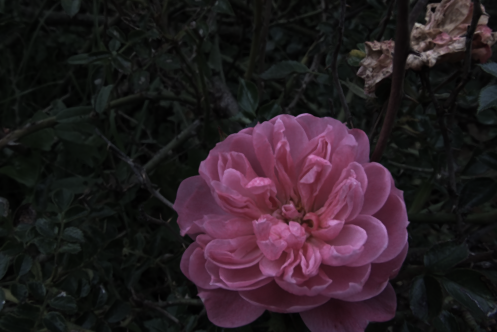}};
  \begin{scope}[x={(image.south east)},y={(image.north west)}]
        \draw[red,thick] (0.78,0.8) rectangle (0.88,0.9);
    \end{scope} 
  \end{tikzpicture}\hfill
\begin{tikzpicture}
  \node[anchor=south west,inner sep=0] (image) at (0,0)
  {\includegraphics[width=0.19\linewidth]{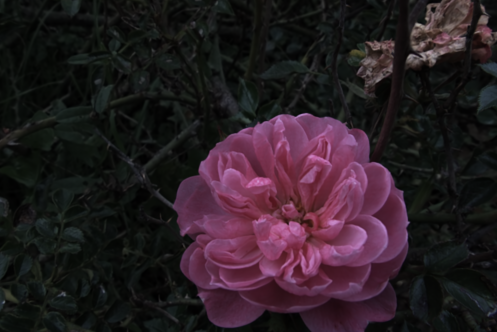}};
  \begin{scope}[x={(image.south east)},y={(image.north west)}]
        \draw[red,thick] (0.78,0.8) rectangle (0.88,0.9);
    \end{scope} 
  \end{tikzpicture}\hfill
\begin{tikzpicture}
  \node[anchor=south west,inner sep=0] (image) at (0,0)
  {\includegraphics[width=0.19\linewidth]{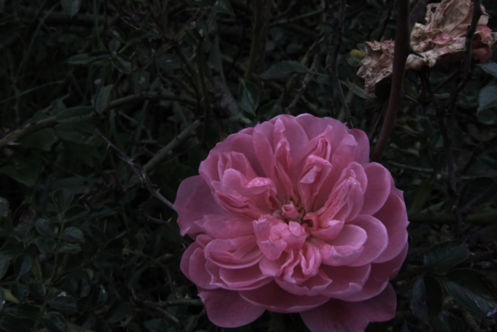}};
  \begin{scope}[x={(image.south east)},y={(image.north west)}]
        \draw[red,thick] (0.78,0.8) rectangle (0.88,0.9);
    \end{scope} 
  \end{tikzpicture}\hfill
\begin{tikzpicture}
  \node[anchor=south west,inner sep=0] (image) at (0,0)
  {\includegraphics[width=0.19\linewidth]{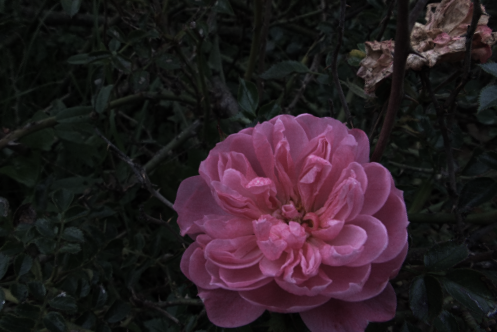}};
  \begin{scope}[x={(image.south east)},y={(image.north west)}]
        \draw[red,thick] (0.78,0.8) rectangle (0.88,0.9);
    \end{scope} 
  \end{tikzpicture}

    \subfloat[40.07 dB\label{fig:test21}]
  {\includegraphics[trim=380 260 60 33, clip,width=0.19\linewidth]{Images/kalantari2.png}}\hfill
  \subfloat[42.74 dB\label{fig:test22}]
  {\includegraphics[trim=380 260 60 33, clip,width=0.19\linewidth]{Images/navarro2.png}}\hfill
  \subfloat[42.94 dB\label{fig:test23}]
  {\includegraphics[trim=380 260 60 33, clip,width=0.19\linewidth]{Images/proposed_wa2.png}}\hfill
  \subfloat[43.38 dB\label{fig:test24}]
  {\includegraphics[trim=380 260 60 33, clip,width=0.19\linewidth]{Images/proposed2.png}}\hfill
  \subfloat[\label{fig:test25}]
  {\includegraphics[trim=380 260 60 33, clip,width=0.19\linewidth]{Images/7824.png}}
\setlength{\belowcaptionskip}{-15pt} 
\caption{Visual comparison of the central view of the scenes. First and second row contains the \textsc{LEAVES} scene and its zoomed-in region from \textit{Diverse} dataset, whereas, third and fourth row contains the \textsc{IMG\_7824} scene and its zoomed-in region from \textit{Flowers} dataset. (a)(f) Kalantari \textit{et al.} \cite{kalantari2016learning}, (b)(g) Navarro \textit{et al.} \cite{navarro2019learning}, (c)(h) LFVS, (d)(i) LFVS-AM, and (e)(j) Ground-truth.}
\label{test123}
\end{figure}
\subsection{Comparison with the state-of-the-art}
To evaluate the performance of the proposed method, named LFVS-AM (Light-Field View Synthesis using Attention Module), we compare it with state-of-the-art methods including Kalantari \textit{et al.} \cite{kalantari2016learning} and Navarro \textit{et al.} \cite{navarro2019learning}. For quantitative evaluation, we calculate the PSNR and MS-SSIM (dB) = $-10\log_{10}(1 - $MS-SSIM$)$ values for the luminance channel of the images between the estimation and the ground-truth. Table \ref{psnr} shows the PSNR and MS-SSIM results averaged over all intermediate target positions -- 7$\times$7 views excluding the 4 corner views which served as input. It is evident that the proposed method produces better results compared to both state-of-the-art methods across all metrics on both the datasets. Our method outperforms \cite{kalantari2016learning} by 2.3 dB and \cite{navarro2019learning} by 0.5 dB on average. Moreover, visual results in Fig. \ref{test123} demonstrates that Kalantari \textit{et al.} \cite{kalantari2016learning} and Navarro \textit{et al.} \cite{navarro2019learning} failed to generate plausible results in the presence of occlusions. For example in Fig. \ref{fig:test11} and \ref{fig:test12}, both of these methods are not able to generate the pole between the two leaves, which is occluded in the input views. In comparison, our method better estimates the occluded region which is closer to the ground-truth, as shown in Fig.~\ref{fig:test14}. 
\subsection{Ablation Study}
\subsubsection{Effects of hyperparameters}
In Table \ref{ablation_param}, we compare three variants of LFVS-AM models having a different number of \textit{ResGroups} and CBAMs. The metrics in the table are calculated for Y channel of the image. From the results, we can conclude that the model with 5 \textit{ResGroup} and 3 CBAM performs the best among the three models, but the differences in performance are not that significant when the total number of CBAM blocks becomes the same i.e., (\# \textit{ResGroups})$\times$(\# CBAMs)=15.
\begin{table}[t!]
\centering
\footnotesize
\setlength\tabcolsep{8pt}
\begin{tabular}{c|c|cc}
\# ResGroups & \# CBAMs & PSNR (dB)  & MS-SSIM (dB) \\ \hline
3          & 3      & 41.16 & 25.76      \\ 
3          & 5      & 41.25 & 25.95      \\
5          & 3      & \textbf{41.33} & \textbf{26.05}      \\
\end{tabular}
\setlength{\belowcaptionskip}{-10pt} 
\caption{Quantitative results for three different models with varying number of \textit{ResGroups} and CBAMs on \textit{Diverse} \cite{kalantari2016learning} dataset.}
\label{ablation_param}
\end{table}
\subsubsection{Effect of the attention mechanism}
To evaluate the effect of the attention mechanism in the refinement network, we compare the proposed method with the model having no attention mechanism. For the model without attention mechanism (LFVS), we remove the CA and SA modules in Fig. \ref{CBAM} (a) to change the CBAM to a residual block. In Table \ref{psnr}, average PSNR and MS-SSIM (dB) values on all intermediate rendering positions are presented. The results indicate the effectiveness of the attention mechanism in our proposed refinement network. Our LFVS model without attention mechanism is better than SoTA because of the residual connections in the network but still can not handle occlusions, as shown in Fig. \ref{fig:test13}. In contrast, the attention module guides the network to efficiently predict the occluded regions, see Fig. \ref{fig:test14}.
\section{Conclusion}
\label{conc}
In this work, we proposed a novel end-to-end CBAM light-field view synthesis method to reconstruct dense light-fields from sparse input data. Our model consists of stereo feature extraction, disparity estimation, and residual CBAM image refinement networks. All these components are modelled using three sequential CNNs. We presented quantitative and qualitative results of the proposed method on two different datasets. Experimental results show that our model outperforms state-of-the-art approaches by 0.5 dB on average. Future work will include to analyze the potential of our proposed method in the context of light-field compression frameworks.
\section{Acknowledgment}
We thank Milan Stepanov (L2S, CentraleSupélec) for his valuable help during the re-implementation of \cite{navarro2019learning}. This project has received funding from the European Union’s Horizon 2020 research and innovation programme under the Marie Skłodowska-Curie grant agreement No. 765911 (RealVision).  

\bibliographystyle{IEEEbib}
{\footnotesize\bibliography{refs}}

\end{document}